\documentstyle[12pt,aaspp4]{article}

\input epsf

\newcommand{\Tb}{T_{\scriptscriptstyle {\rm b}}}
\newcommand{\Tbnu}{T_{{\scriptscriptstyle {\rm b}},\nu}}
\newcommand{\tauT}{\tau_{\scriptscriptstyle {\rm T}}}
\newcommand{\nuo}{\nu_{\scriptscriptstyle 0}}
\newcommand{\no}{n_{\scriptscriptstyle 0}}

\newcommand{\nH}{n_{\scriptscriptstyle {\rm H}}}

\newcommand{\gtapprox}{\raisebox{-0.5ex}{$\,\stackrel{>}{\scriptstyle
\sim}\,$}}
\newcommand{\ltapprox}{\raisebox{-0.5ex}{$\,\stackrel{<}{\scriptstyle
\sim}\,$}}

\received{}
\revised{}
\accepted{}

\lefthead{Kuncic, Bicknell \& Dopita}
\righthead{Induced Scattering in GPS Sources}

\begin{document}

\title{Induced Compton Scattering in Gigahertz Peak Spectrum Radio Sources}
\author{Zdenka Kuncic\altaffilmark{1}, Geoffrey V. Bicknell}
\affil{Astrophysical Theory Centre\altaffilmark{2}, \\
Australian National University, Canberra, ACT 0200, Australia}
\altaffiltext{1}{Now at Dept. Physics \& Astronomy, University of Victoria,
B.C., V8W 3P6, Canada. Email: {\tt zdenka@uvastro.phys.uvic.ca}.}
\altaffiltext{2}{The ANUATC is jointly operated by the Mount Stromlo and
Siding Spring Observatories and the School of Mathematical Sciences.}
\and
\author{Michael A. Dopita}
\affil{Mount Stromlo and Siding Spring Observatories, Weston PO, ACT 2611,
Australia}

\begin{abstract}
We revisit the shocked shell model for the class of Active Galactic
Nuclei known as Gigahertz Peak Spectrum sources, incorporating
new observational data on the radiation brightness temperatures.
We argue that in addition to free--free absorption, induced Compton
scattering will also have an important effect in forming the $\sim$~GHz
peak and in shaping the radio spectra that characterize these sources.
Indeed, our arguments suggest that GPS sources may provide the first real
evidence for the role of induced Compton scattering in extragalactic radio
sources.
\end{abstract}

\keywords{galaxies: active --- shocks ---
radiation mechanisms: thermal, non--thermal --- scattering}

\section{Introduction}

Recently, Bicknell, Dopita, \& O'Dea (1997) developed a model which unifies
the Gigahertz Peak Spectrum (GPS) sources and the related Active Galactic
Nuclei (AGN) classes known as Compact Steep Spectrum (CSS) sources and Compact
Symmetric Objects (CSOs).
Their model attributes the observed radio spectra and optical emission to the
interaction of a jet--driven, non--thermal lobe with the ambient galactic
interstellar medium (ISM).
As the radio lobe forces its way through the ISM, strong radiative shocks
create shells of shock--ionized and photo--ionized gas which
produce a low--frequency break in the power law spectrum through free--free
absorption (FFA). Although synchrotron self--absorption (SSA) provides a
natural mechanism to produce a low frequency break in a non--thermal spectrum
and is successful in explaining the spectra of AGN with distinct core--jet
morphologies (\cite{BlandKon79}), we expect that it is unimportant in the more
extended jet lobes of GPS sources, where the non--thermal plasma is more
likely to be optically--thin to its own radiation.
However, the spectrum emitted by the non--thermal lobe plasma can still
exhibit a low frequency break not only as a result of FFA in an external
shell of thermal gas surrounding the lobe, as suggested by Bicknell et al.
(1997), but also as a result of the competing low--frequency process of
induced Compton scattering (ICS).

ICS becomes important in a high brightness temperature radiation field, such
as that which is readily generated by synchrotron radio sources and which has
a Thomson scattering optical depth $\tauT$ and a brightness temperature
$k\Tbnu \gg m_{\rm e}c^2 \gg h\nu$, such that
\begin{equation}
\left(  f \frac{k\Tbnu}{m_{\rm e}c^2} \right) \tauT \gtapprox 1 \; ,
\label{eq:opdepth}
\end{equation}
where the parameter $f$ quantifies the strong effect of the
angular distribution of the radiation and is given by:
\begin{equation}
f \approx \frac {3}{16\pi} \int (1+\cos^2 \phi) (1-\cos \phi) d\Omega
\label{eq:f}
\end{equation}
where $\phi$ is the scattering angle and
the integral is over the solid angle of the beam incident on the
scattering region (\cite{CopBlandRees93}). For a conical beam of radiation
with half angle $\cos^-1(\mu_0)$ and solid angle $\Omega_0$,
$f = (1-\mu_0)^2(3\mu_0^2+2\mu_0+7)/32 $; for complete isotropy of the
radiation field $f = 1$; for $\mu_0$ near unity, $f\approx
\frac{3}{8}(1-\mu_0)^2$.
In both cases $f\approx (\Omega_0/4\pi)^2$.
ICS  redistributes photon energies from frequencies near where $\Tbnu$ peaks
to lower frequencies where they are absorbed (e.g. by SSA or FFA) effectively
reducing the peak $\Tbnu$ observed. As first demonstrated by
\cite{Sunyaev70}, the key observational consequence is a break in synchrotron
radio spectra.

Equation (\ref{eq:opdepth}) implies that induced scattering can produce
spectral distortions when $\tauT \gtapprox (5\times 10^9){\rm K}/f\Tb$ and
hence, unlike `ordinary' (spontaneous) Compton scattering, it can be
important even when $\tauT < 1$.
Indeed, for a typical synchrotron radio source, ICS can be more efficient
than SSA in producing a spectral turnover and can limit the brightness
temperature at GHz frequencies to
$T_{\rm b,1 GHz} \ltapprox 10^{11} \nu_9^{-0.2}
\gamma_{\rm min}^{0.6}$~K, where $\gamma_{\rm min}$ is the low--energy
cutoff to the non--thermal electron distribution\footnote{A characteristic
power--law index of 2.4 has been assumed here for the electron distribution.
According to standard synchrotron theory, this corresponds to a spectral index
of $\alpha \simeq 0.7$, which is a fiducial value for the GPS sources we are
concerned with here (Bicknell et al. 1997).} (\cite{SincKrol94}).
Note that this limit is only weakly dependent on the source parameters
and (for $f\sim 1$) is lower than the theoretical inverse Compton limit,
$\Tb \ltapprox 10^{12}$~K, for typical AGN sources (\cite{HoyBurbSarg66}).
As argued by Sincell \& Krolik (1994), ICS may explain
why the inverse Compton limit is, in practice, rarely observed.
Such is the case for GPS sources, where the observed brightness temperatures
of lobe subcomponents are typically $10^{9-10}$~K (\cite{Stang97}).
Since ICS by the non--thermal, relativistic electrons in the jet lobes cannot
reduce the intrinsic peak $\Tb$ to these observed levels, then ICS may
instead be operating externally.
Indeed, Coppi, Blandford \& Rees (1993) have shown that ICS by relatively
`cold' (nonrelativistic) electrons outside of a primary non--thermal source can
effectively reduce the peak brightness temperature to values such that
$k\Tb \simeq m_{\rm e}c^2$ and this alone makes it very tempting to adopt
such a scenario to explain the observed $\Tb$ values in GPS sources.

In this {\em Letter}, we examine the process of induced scattering to
show that in order to account for the observed brightness temperatures
in GPS sources, this process must be operating at some level in the
shock--ionized shell that has been postulated to surround the non--thermal
lobes.

\section{The Case for Induced Scattering in GPS Sources}

In the scenario proposed by \cite{BickDopOdea97}, the non--thermal jet
lobes in GPS sources are surrounded by a shell of thermal, nonrelativistic
plasma formed from shocked and photoionized material in the ambient ISM.
Although FFA is likely to be operating at some level in this external shell,
there will be strong competition from ICS, especially if the shell is tenuous.
Indeed, with the expected brightness temperatures, $\sim 10^{11}$~K, of the
primary non--thermal radiation which impinges upon the shell, ICS is more
efficient than FFA in producing a break at $\sim$ GHz frequencies when the
electron number density, $n_{\rm e}$, satisfies
\begin{equation}
n_{\rm e} \ll 100 \, \nu_9^2 \, T_4^{3/2}
\left( \frac{T_{\rm b,1 GHz}}{10^{11}{\rm K}} \right) \, {\rm cm}^{-3} \, ,
\label{eq:density}
\end{equation}
where the electron temperature, $T_{\rm e} = 10^4 T_4$~K.
This is precisely the range of densities relevant to shock--ionized
ISM material surrounding jet lobes in GPS sources, as found by
\cite{BickDopOdea97}.
Although, in general, collective plasma effects will not have an overall
affect on ICS (see \cite{CopBlandRees93}), this is not strictly the case
in the presence of a high brightness temperature radiation field, which can
trigger stimulated Raman scattering, whereby the radio waves scatter off
plasmons (Langmuir waves) rather than off the individual electrons themselves.
Since, at the densities relevant here, this process requires brightness
temperatures $\gtapprox 10^{14}$ K  (\cite{LevBland95}), it is unlikely
to be important in this context.

A definitive test for any model which attempts to explain the spectral peak
 in GPS/CSS sources is the observed inverse correlation between the frequency
at which the spectrum peaks, $\nu_{\rm p}$, and the source size. In the FFA
model of \cite{BickDopOdea97}, a theoretical relationship between these two
parameters readily emerges from the shock dynamics. We follow the same method
here in constructing an analogous ICS model, but as we explain below, the
predicted relationship has different dependences on the physical parameters.

A spectral break due to ICS is expected at the frequency where the optical
depth, given by the left hand side of equation (\ref{eq:opdepth}), reaches
unity. However, as equations~(\ref{eq:opdepth}) and  (\ref{eq:f}) show,
induced scattering depends strongly on the angular distribution of the high
brightness temperature radiation field and we have allowed for this through the
parameter $f$.
The flux density corresponding to the condition
on the brightness temperature is obtained by simply
integrating the surface brightness over the solid angle $\Omega$ subtended
by the source at the observer, giving
\begin{equation}
\nu^{-2} F_\nu = 2 m_{\rm e} \int (f \, \tauT)^{-1} \> d\Omega \, .
\label{eq:flux}
\end{equation}
In the model developed by \cite{BickDopOdea97}, $\tauT$ is approximately
uniform over the sides of the lobe and the solid angle can be expressed as
$\Delta \Omega \simeq \pi r_{\rm c} x_{\rm h}/D^2=\pi
\zeta^{-1/2} x_{\rm h}^2 /D^2$,
where $r_{\rm c}$ is the radius of the semi--ellipsoidal cocoon which forms
as the jet--driven lobe expands into the ISM, $x_{\rm h}$ is the distance
from the core at the base of the jet to the lobe hotspot (i.e. the semi--major
axis of the cocoon), $D$ is the distance to the source and  $\zeta (\simeq 2$
typically) is the ratio of the averaged hotspot pressure to cocoon pressure.
By taking a power--law spectrum, $F_{\nu} = F_{\nuo}(\nu / \nuo)^{-\alpha}$,
where $\nuo$ is a fiducial frequency (5 GHz in this case), and using the
monochromatic power $P_{\nuo} = 4\pi D^2 F_{\nuo}$, equation (\ref{eq:flux})
predicts a peak frequency given by
\begin{equation}
\frac {\nu_{\rm p}}{\nuo} \simeq
\left[ \frac {f \, \zeta^{1/2} P_{\nuo} \tauT}
{8 \pi^2 x_{\rm h}^2 \nuo^{2} m_{\rm e} }\right]^{\frac {1}{\alpha+2}}\;,
\label{eq:nupeak}
\end{equation}
where $P_{\nuo} = 4\pi D^2 F_{\nuo}$ is the monochromatic power.

The Thomson optical depth is assumed to be approximately uniform over the
sides of the lobe and  is calculated with the {\small MAPPINGSII} code
(\cite{SuthDop93}) for radiative shocks, giving
\begin{equation}
\tauT \simeq (4.9 \times 10^{-3}) \, V_3^{3.6} +
(1.6 \times 10^{-2}) \, V_3^{2.5} \, ,
\label{eq:tauT}
\end{equation}
where $V = 1000V_3 \, {\rm km \, s}^{-1}$ is the shock velocity corresponding
to the sideways expansion of the radio lobe.
The two terms correspond to the shocked gas and photoionized precursor.
Specifically, the shock velocity is given by (see equations 2-10 and 2-11 in
Bicknell et al. 1997)
\begin{equation}
V_3 = V_{0,3} \, \left( \frac {x_{\rm h}}{\rm kpc}
\right)^{\frac{\delta - 2}{3}} \, ,
\label{eq:vel_scale}
\end{equation}
where
\begin{equation}
V_{0,3} = 1.5 \> \left( \frac{6}{8-\delta} \right)^{1/3} \:
\zeta^{1/6} \:
\left[ \frac {F_{\rm E,45}}{\no} \right]^{1/3}
\label{eq:velocity}
\end{equation}
and where $F_{\rm E} = 10^{45} F_{\rm E,45} \, \rm erg \, s^{-1}$ is the jet
energy flux and $\delta$ is the power--law index of the Hydrogen density
distribution, given by $\nH = \no (x_{\rm h} / {\rm kpc})^{-\delta}$.

Equation (\ref{eq:nupeak}) predicts a relationship between peak frequency and
source size that derives quite differently from that predicted by the FFA model
(see equation~5-10 in \cite{BickDopOdea97}). Since the opacity of ICS is
proportional to the radiation brightness temperature, the chief dependence of
$\nu_{\rm p}$ derives from the integrated surface brightness ($\propto x_{\rm
h}^2$). There is also a weaker dependence on $x_{\rm h}$ through the shock
velocity, as given in equation (\ref{eq:vel_scale}), which in turn determines
$\tauT$. In the case of the FFA model, on the other hand, while there is also a
similar weak dependence through $V$, the predicted $\nu_{\rm p}$~--~$x_{\rm h}$
relationship derives chiefly from the dependence on density (which has a
power--law scaling with source size). This density dependence arises simply
because the opacity due to FFA is
$\propto \int n_{\rm e}^2 dl$, whereas for ICS, it is
$\propto \int n_{\rm e} dl$ where both integrals are through the ionized
screen.
Allowing for the inverse dependence of the shocked cooling zone and the
photoionized precursor on density, the dependences on density are respectively
on the first and  zeroth powers of $n_e$. This is reflected, for instance, in
the dependence of the Thomson optical depth on velocity alone and this is where
a dependence on density does enter, since the expansion velocity is a
function of
the ambient density. Although these two different models predict a $\nu_{\rm
p}$~--~$x_{\rm h}$ relationship that derives from intrinsically different
physical processes (i.e. two--body interactions versus particle--photon
interactions), the overall differences are surprisingly small: both FFA and ICS
models predict a negative slope in the $\nu_{\rm p}$~--~$x_{\rm h}$ plane which
is close to unity and which varies very weakly with the relevant range of
parameters.

Let us now examine the predicted peak frequency - source size relationship
more quantitatively.

\section{Results}

The relationship between peak frequency, $\nu_{\rm p}$, and source size,
$x_{\rm h}$, as predicted by equation (\ref{eq:nupeak}) for the anisotropy
parameter $f=1$ is plotted and
compared with observational data (\cite{Fanti90,Stang97,OdeaBaum97}) in
figure~1.
The parameters involved in the theoretical lines are as follows: values of
the Hydrogen densities at 1 kpc, $\no$, are 1 and $10 \> \rm cm^{-3}$; values
of the density power--law index, $\delta$, are 1.5 and 2 (these are the lower
and upper limits suggested by \cite{Begelman96} in his fit
to the luminosity--size relation for CSOs); the total 5 GHz radio power is
$P_{\rm 5 GHz} = 10^{27.5} \> \rm erg \; s^{-1} \: Hz^{-1}$, corresponding
to the average radio power of GPS sources (\cite{Fanti90,Stang97}); the jet
energy flux is $F_{\rm E} = 10^{45} \, \rm ergs \, s^{-1}$ and the spectral
index is $\alpha = 0.7$ (Bicknell et al. 1997).
Table~1 lists the values of the parameter $V_{0,3}$, which provides the scale
for the size of the velocity as a function of source size
(see eq.~[\ref{eq:velocity}]).

The slopes of both the $\delta=1.5$ and $\delta=2$ lines are adequate fits to
the slope of the data points and the overall fits to the data are obviously
better for the lower values of $\no$. The fits could also be improved by
increasing the radio power and Thomson optical depth. Increasing the radio
power increases the brightness temperature of the lobe and hence the frequency
at which induced Compton scattering becomes important. We have not decreased
the density still further to produce a more acceptable fit since it is evident
from table~1 that lower densities would involve an unacceptable extrapolation
of the fit to the Thomson opacity, well beyond the velocities of $\sim 1000 \>
\rm km \> s^{-1}$ that the {\small MAPPINGSII} code can adequately handle.
Indeed, the velocity in even the $\delta=1.5$, $\no=10$ model is greater than
$1000 \> \rm km \: s^{-1}$ for an overall size $\ltapprox 0.5 \rm kpc$. It can
be reasonably assumed, however, that the lower densities favoured by our ICS
model imply cocoon velocities in excess of $1000 \> \rm km \: s^{-1}$, even if
the extrapolation of the shock modeling beyond such a velocity is uncertain.

\section{Discussion}

Our model assumes a uniform distribution of radio emitting plasma in the
jet lobes,
whereas in a number of imaged GPS sources,  the emitting region is clumpy;
the highest
brightness temperatures are typically identified with a small number of
localized
components (\cite{Stang97,conway94a,wilkinson94a}). If these clumps were
embedded in a
lower surface brightness cocoon outside of which the ionizing screen were
located then
the radiation intercepting this screen would be anisotropic.   This  effect of
anisotropy is countered however, by the power (typically
$1/2.7$)  entering in the expression for $\nu_p$.  Thus, $\nu_p$ would be
reduced to
approximately 40\% of the isotropic value  for $f\sim 0.1$. Moreover, as
inspection of
the images of GPS and CSO sources (the latter all being GPS) reveals, the
"clumps" are
usually situated at increasing distances from the nucleus with a local
filling factor
of order unity. The beam solid angle corresponding to each clump should
therefore be
of order $2 \pi$. Anisotropy need not be an  important factor especially in
view of the insensitivity of the peak frequency to it.

On the other hand,  non-uniformity in the emission increases the brightness
temperature. In our uniform model, the brightness Temeprature at 5~GHz, is
given by:
\begin{eqnarray}
< T_{\rm b,5 GHz} > \; \simeq \frac{\zeta^{1/2} c^2  P_\nu}
{8 \pi^2 x_{\rm h}^2 \nu^2 k} \simeq 10^{10} \>
\left( \frac{P_{\rm 5 GHz}}{10^{27.5} \, \rm W \: Hz^{-1}} \right)
\, \left( \frac{x_{\rm h}}{100 {\rm pc}} \right)^{-2}
\left( \frac{\nu}{5 \times 10^9 {\rm Hz}} \right) ^{-2} \, {\rm K} \, .
\end{eqnarray}
and is about an order of magnitude lower than the values observed in
subcomponents by \cite{Stang97}, for example.
A larger brightness temperature of course {\em enhances} the prospect of
induced Compton scattering.

Interestingly, the Stanghellini et al. study of a complete sample of GPS
sources has revealed a pronounced peak in the distribution of low--frequency
spectral indices near $\alpha = -1.0$, which is in remarkabe agreement with
the theoretically predicted spectra resulting from induced Compton scattering
by external thermal electrons (e.g. \cite{CopBlandRees93}).
This feature, combined with the successful ICS explanation for the peak
frequency -- source size relationship, shows that GPS sources may indeed
constitute the first real instance where induced Compton scattering plays a
relatively clear role in extragalactic radio sources.

We therefore conclude that induced Compton scattering may be an important
process in the shocked and photoionized plasma surrounding an expanding lobe
in GPS sources; it can contribute to the formation of the characteristic peak
in the radio spectrum when the density falls below $\sim 10 \> \rm cm^{-3}$
and free--free absorption becomes inefficient.
Expansion velocities of $1000 - 2000 \> \rm km \: s^{-1}$ are then implied
when induced scattering is the dominant process.
Induced scattering may thus be particularly relevant in the high redshift GPS
sources (\cite{Elvis97}), which have column depths of the order of a few
$\times 10^{22}$ cm$^{-2}$, at the lower end of the range estimated by
\cite{BickDopOdea97} on the basis of the FFA model.

In general, the introduction of induced Compton scattering not only increases
the range of plasma densities over which a low frequency break can be produced,
but also increases the efficiency of FFA at the lowest frequencies, where
photons are being continuously replenished as a result of the downscattering
of GHz photons.
Indeed, the combined effect of induced scattering and free-free
absorption could also be important in some other sources, such as NGC 1275,
in which one of the jets appears to be significantly weaker than the other
at low frequencies.
This has been attributed to free--free absorption (\cite{LevLaorVerm95}) and
although the observations are on parsec scales in this particular source,
the brightness temperatures are similar to the values considered here,
suggesting that induced Compton scattering may well be operating on various
jet scales, at the very least in conjuction with the more conventional process
of free--free absorption.

\acknowledgments{We wish to thank Professor Rashid Sunyaev for suggesting that
we consider ICS for GPS sources and our anonymous referee whose valuable
comments and suggestions helped to improve the paper significantly.
We also thank Chris O'Dea for helpful comments.}

\newpage

\section*{Figure Caption}

\begin{figure}
\centerline{\epsfbox{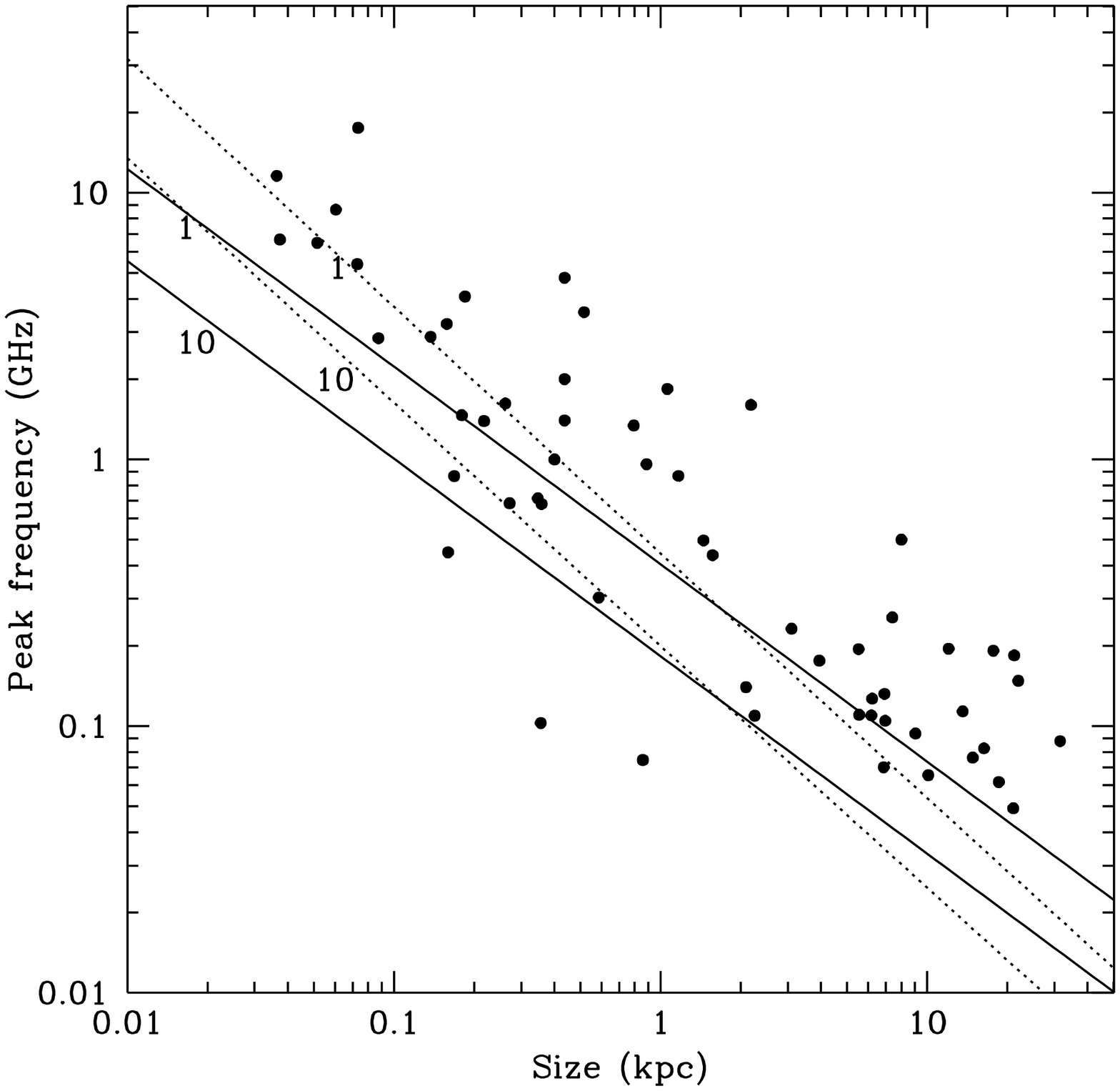}}
\end{figure}

\figcaption{A comparison of the predicted and observed relationship
between the peak frequency and source size for GPS sources.
The lines represent the theoretical relationship predicted by equation~
(\ref{eq:nupeak}): solid and dashed lines are for $\delta = 2.0$ and
$\delta = 1.5$, respectively (values of $\no =1 \,{\rm cm}^{-3}$ and
$\no = 10 \,{\rm cm}^{-3}$ are indicated) and the other values used are
$\alpha = 0.7$, $P_{\rm 5 GHz} = 10^{27.5} \,{\rm W \, Hz}^{-1}$ and
$F_{\rm E} = 10^{45} {\rm erg \, s}^{-1}$ (see the text for
definitions of these parameters). The data points are from Fanti et~al. (1990),
O'Dea \& Baum (1997)and Stanghellini et~al. (1997).}

\newpage

\begin{center}
{\bf Table 1: Parameters of expanding cocoon}
\vskip 11 pt

\begin{tabular}{r r r}
\hline
\multicolumn{1}{c}{$\no$}         & \multicolumn{1}{c}{$\delta$} &
\multicolumn{1}{c}{$V_{0,3}$}  \\
\multicolumn{1}{c}{$\rm cm^{-3}$} &                              &
                               \\
\hline
1  & 2   & 1.80 \\
10 & 2   & 0.84 \\
1  & 1.5 & 1.75 \\
10  & 1.5 & 0.81 \\
\end{tabular}
\end{center}

\end{document}